# Genesis of general relativity – a concise exposition*[§]

Wei-Tou Ni
*School of Optical-Electrical and Computer Engineering,
University of Shanghai for Science and Technology,
516, Jun Gong Rd., Shanghai 200093, China* weitou@gmail.com

This short exposition starts with a brief discussion of situation before the completion of special relativity (Le Verrier's discovery of the Mercury perihelion advance anomaly, Michelson-Morley experiment, Eötvös experiment, Newcomb's improved observation of Mercury perihelion advance, the proposals of various new gravity theories and the development of tensor analysis and differential geometry) and accounts for the main conceptual developments leading to the completion of the general relativity: gravity has finite velocity of propagation; energy also gravitates; Einstein proposed his equivalence principle and deduced the gravitational redshift; Minkowski formulated the special relativity in 4-dimantional spacetime and derived the 4-dimensional electromagnetic stress-energy tensor; Einstein derived the gravitational deflection from his equivalence principle; Laue extended the Minkowski's method of constructing electromagnetic stress-energy tensor to stressed bodies, dust and relativistic fluids; Abraham, Einstein, and Nordström proposed their versions of scalar theories of gravity in 1911-13; Einstein and Grossmann first used metric as the basic gravitational entity and proposed a "tensor" theory of gravity (the "Entwurf" theory, 1913); Einstein proposed a theory of gravity with Ricci tensor proportional to stress-energy tensor (1915); Einstein, based on 1913 Besso-Einstein collaboration, correctly derived the relativistic perihelion advance formula of his new theory which agreed with observation (1915); Hilbert discovered the Lagrangian for electromagnetic stress-energy tensor and the Lagrangian for the gravitational field (1915), and stated the Hilbert variational principle; Einstein equation of general relativity was proposed (1915); Einstein published his foundation paper (1916). Subsequent developments and applications in the next two years included Schwarzschild solution (1916), gravitational waves and the quadrupole formula of gravitational radiation (1916, 1918), cosmology and the proposal of cosmological constant (1917), De Sitter solution (1917), Lense-Thirring effect (1918).

---

## 1. Prelude – Before 1905

General Relativity (GR) was fast in its acceptance in the world community. This was not the case for Newtonian gravitation [1]. We quote from the beginning of Chapter V on Gravitation of Vol. II from Whittaker [2]: "We have seen (cf. Vol. I, pp. 29-31) that for many years after its first publication, the Newtonian doctrine of gravitation was not well received. Even in Newton's own University of Cambridge, the textbook of physics in general use during the first quarter of the eighteen century was still Cartesian: while all the great mathematicians of the Continent – Huygens in Holland, Leibnitz in Germany, Johann Bernoulli in Switzerland, Cassini in France – rejected the Newtonian theory altogether."

"This must not be set down entirely to prejudice: many well-informed astronomers believed, apparently with good reason, that the Newtonian law was not reconcilable with the observed motions of the heavenly bodies. They admittedly that it explained satisfactorily the first approximation to the planetary orbit, namely that they are ellipses with the sun in one focus: but by the end of seventeenth century much was known observationally about the departures from elliptic motion, or *inequalities* as they are called, which were presumably due to mutual gravitational interaction: and some of these seemed to resist every attempt to explain them as consequences of the Newtonian law".

The most serious one was the *Great inequality of Jupiter and Saturn*. In the same page, Whittaker continued: *"*A comparison of the ancient observations cited by Ptolemy in the Almagest with those of the earlier astronomers of Western Europe and their more recent successors, showed that for centuries past the mean motion, or average angular velocity round the sun, of Jupiter, had been continually increasing, while the mean motion of Saturn had been continually decreasing." According to Kepler's [3] third law, the orbit of Jupiter must be shrinking and the orbit of Saturn must be expanding. This stimulates the development of celestial mechanics. Euler and Lagrange made significant advances. In 1784, Laplace found that the *Great inequality* is not a secular inequality but a periodic inequality of 929-year long period due to nearly commeasurable orbital periods of Jupiter and Saturn. Calculation agreed with observations. The issue was completely solved. For a more thorough study of the history of the *Great inequality of Jupiter and Saturn*, see the doctoral thesis of Curtis Wilson [4].

In 1781, Herschel discovered the planet Uranus. Over years, Uranus persistently wandered away from its expected Newtonian path. In 1834, Hussey suggested that the deviation is due to perturbation of an undiscovered planet. In 1846, Le Verrier predicted the position of this new planet. On September 25, 1846, Galle and d'Arrest found the new planet, Neptune, within one degree of arc of Le Verrier's calculation. This symbolized the great achievement of Newton's theory. [5]

With the discovery of Neptune, Newton's theory of gravitation was at its peak. As the orbit determination of Mercury reached $10^{-8}$, relativistic effect of gravity showed up. In 1859, Le Verrier discovered the anomalous perihelion advance of Mercury [6].

*Anomalous perihelion advance of Mercury.* In 1840, Arago suggested to Le Verrier to work on the subject of Mercury's motion. Le Verrier published a provisional theory in 1843. It was tested at the 1848 transit of Mercury and there was not close agreement. As to the cause, Le Verrier [7] wrote "Unfortunately, the consequences of the principle of gravitation have not be deduced in many particulars with a sufficient rigour: we will not be able to decide, when faced with a disagreement between observation and theory, whether this results completely from analytical errors or whether it is due in part to the imperfection of our knowledge of celestial physics." [7, 8]

In 1859, Le Verrier [6] published a more sophisticated theory of Mercury's motion. This theory was sufficiently rigorous for any disagreement with observation to be taken quite confidently as indicating a new scientific fact. In this paper, he used two sets of observations --- a series of 397 meridian observations of Mercury taken at the Paris Observatory between 1801 and 1842, and a set of observations of 14 transits of Mercury. The transit data are more precise and the uncertainty is of the order of 1". The calculated planetary perturbations of Mercury are listed in Table 1 [6, 8]. In addition to these perturbations, there is a 5025"/century general precession in the observational data due to the precession of equinox. The fit of observational data with theoretical calculations has discrepancies. These discrepancies turned out to be due to relativistic-gravity effects. Le Verrier attributed these discrepancies to an additional 38" per century anomalous advance in the perihelion of Mercury. [7]

Table 1. Planetary perturbations of the perihelion of Mercury [6, 8].

| | |
|---|---|
| Venus | 280".6/century |
| Earth | 83".6/century |
| Mars | 2".6/century |
| Jupiter | 152".6/century |
| Saturn | 7".2/century |
| Uranus | 0".1/century |
| Total | 526".7/century |

Newcomb [9] in 1882, with improved calculations and data set, obtained 42".95 per century anomalous perihelion advance of Mercury. The value more recently (1990) was (42".98 ± 0.04)/century [10]. At present, ephemeris fitting reached $10^{-4}$ precision. See Ref. [11] and references therein.

*Michelson-Morley experiment.* According to Newton's second law of motion and Galilean transformation, light velocity would change in a moving frame. However, this is not the experimental finding of Michelson and Morley in 1887 [12]: "Considering the motion of the earth in its orbit only, this displacement should be $2Dv^2/V^2 = 2D \times 10^{-8}$. The distance D was about eleven meters, or $2 \times 10^7$ wavelengths of yellow light; hence the displacement to be expected was 0.4 fringe. The actual displacement was certainly less than the twentieth part of this kind, and probably less than the fortieth part. But since the displacement is proportional to the square of the velocity, the relative velocity of the earth and the ether is probably less than one sixth the earth's orbital velocity, and certainly less than one-fourth." D is the optical path length in one arm of the multi-reflection Michelson-Morley interferometer sat on the granite floating in liquid mercury; v is the velocity of earth relative to ether; V is the light velocity. In modern Michelson-Morley experiments, one measures the frequency changes $\Delta v/v$ of two perpendicular Fabry-Perot cavities. The most precise experiment by Nagel *et al.* [13] measured the changes $\Delta v/v$ of two cryogenic cavities to be $(9.2 \pm 10.7) \times 10^{-19}$ (95% confidence interval), a nine order improvement to the original Michelson-Morley experiment.

*Eötvös experiment.* In 1889, Eötvös [14] used a torsion balance with different types of sample materials to significantly improve on the test of the Galileo equivalence principle (the equivalence of gravitational mass and inertial mass; the universality of free fall) [15] to a precision of 1 in 20 million ($5 \times 10^{-8}$). The most recent terrestrial experiments of Washington group used torsion-balance to compare the differential accelerations of beryllium–aluminum and beryllium–titanium test-body pairs with precisions at the part in $10^{13}$ level and confirmed the Galileo equivalence principle [16]. The first space experiment Microscope (MICRO-Satellite à trainée Compensée pour l'Observation du Principe d'Équivalence) [17, 18] has been in orbit since 26 April, 2016 with the aim of improving the test accuracy to one part in $10^{15}$ level and is performing functional tests successfully [18]. The Microscope test masses are made of alloys of Platinum-Rhodium (PtRh10 – 90% Pt, 10% Rh) and Titanium-Aluminum-Vanadium (TA6V – 90% Ti, 6% Al, 4% V), while the REF test masses are made of the same PtRh10 alloy. The weak equivalence for photons are confirmed with precisions at the part in $10^{38}$ level in astrophysical and cosmological observations on electromagnetic wave propagation [19].

*The discovery of Mercury perihelion advance anomaly undermined Newton's gravitation theory while the null results of Michelson and Morley undermined the Galilean invariance and Newton's dynamics.* The foundation of Newton's world system and classical physics needed to be replaced. The precise verification of weak equivalence principle and realization that the phenomena are the same in a uniformly moving boat and on ground made it easier to advance one step in cognition to comprehend and formulate Einstein equivalence principle (the phenomena in a falling elevator are the same as in free space).

In the last half of the 19th century, efforts to account for the anomalous perihelion advance of Mercury explored two general directions: (i) searching for a putative planet 'Vulcan' or other matter inside Mercury's orbit; and (ii) postulating an *ad hoc* modified gravitational force law. Both these directions proved unsuccessful. Proposed modifications of the gravitational law included Clairaut's force law (of the form $A/r^2 + B/r^4$), Hall's hypothesis (that the gravitational attraction is proportional to the inverse of distance to the $(2+\delta)$ power instead of the square), and velocity-dependent force laws. The reader is referred to Ref. [8] for an in-depth history of various proposed theories and their references.

A compelling solution to this problem had to await the development of general relativity. When general relativity is taken as the correct theory for predicting corrections to Newton's theory, we understand why when the observations reached an accuracy of the order of 1" per century (transit observations), a discrepancy would be seen. Over a century, Mercury orbits around the Sun 400 times, amounting to a total angle of $5 \times 10^8$ arcsec. The fractional relativistic correction (perihelion advance anomaly) of Mercury's orbit is of order $\xi G_N M_{Sun}/dc^2$ with $d$ being the distance of Mercury to the Sun and $\xi$ a parameter of order one depending on theory; for general relativity with $\xi = 3$, it is $8 \times 10^{-8}$. Therefore, the general relativistic correction for perihelion advance is about 40 arc sec per century. As the orbit determination of Mercury reached an accuracy of order $10^{-8}$, the relativistic corrections to Newtonian gravity became manifest.

We thus see how gravitational anomalies can lead either to the discovery of missing matter or to a modification of the fundamental theory for gravity. It is not totally coincidental that Le Verrier not only predicted the position of a new planet, but also discovered the Mercury perihelion advance anomaly as astronomical observation were refined and accumulated for a century.

*Michelson-Morley experiment inspired the consideration of new covariant formulation of electromagnetism under reference frame transformation.* Michelson-Morley experiment, with various proposals and developments, led eventually to the approximate transformation theory of Lorentz [20], and the principle of relativity of Poincaré [21-23]. In 1901, Poincaré [25] performed a rigorous mathematical and

physical analysis of various variants of the electrodynamic theory; in the introduction, he wrote (English translation from pp. 48-49 of [26]): "Although none of these theories seems to me fully satisfactory, each one contains without any doubt a part of the truth and comparing them maybe instructive. From all of them, Lorentz theory seems to me the one which describes in the better way the facts." What Poincaré used as a criterion of satisfaction is whether the principle of relative motion is fully satisfied (Ref. 21, p. 477): The movement of any system whatever ought to obey the same laws, whether it is referred to fixed axes or to the movable axes which are implied in uniform motion in a straight line (English translation from p. 63 of [26]). This is clearly the invariance of the laws under a change of reference frame, when one is related to the other by a constant velocity [26]. Nevertheless, in 1900, the transformation from one frame to the other was not known. The "classical" composition law for velocities was clearly not working for explaining optical experiments such as Michelson and Morley's [26].

In 1902, Poincaré called the principle of relative motion as the principle of relativity [22, 23]. In 1904 Poincaré gave a talk entitled *L'état actuel et l'avenir de la physique mathématique* to the scientific congress at the Saint Louis World Fair and stated the Principle of Relativity [23, 24] as "*The laws of physical phenomena must be the same for a fixed observer and for an observer in rectilinear and uniform motion so that we have no possibility of perceiving whether or not we are dragged in such a motion*". In the same year, *Lorentz* [20] *formulated an approximate transformation theory which satisfied the principle of relativity and agreed with all the experiments to their precision at that time.*

*In 1905, using the principle of relativity, Poincaré* [27] *arrived at the exact invariant transformation* (Poincaré called it the Lorentz transformation) *and completed the transformation theory of special relativity.* In a subsequent paper, Einstein [28] also arrived at the exact Lorentz transformation and completed the transformation theory of special relativity. Thus, the special theory of relativity was born. For a more complete study of the history of the development of special theory of relativity, we refer the readers to Messager and Letellier's review [26].

*In addition to the transformation theory of special theory of relativity, Einstein* [29] *made a cognition advance in postulating and ascertaining the general mass-energy equivalence relation: $E = mc^2$.* For a brief history of the genesis of the mass-energy equivalence relation, we quote Whittaker (pp. 51-52 of [2]):

"We have now to trace the gradual emergence of one of the greatest discoveries of the twentieth century, namely, the connection of mass and energy."

"As we have seen,[1] J. J. Thomson in 1881 arrived at the result that a charged spherical conductor moving in a straight line behaves as if it had an additional mass of amount $(4/3c^2)$ times the energy of its electrostatic field.[2] In 1900 Poincaré,[3] referring to the fact that in free aether the electromagnetic momentum is $(1/c^2)$ times the Poynting flux of energy, suggested that electromagnetic energy might possess mass density equal to $(1/c^2)$ times the energy density: that is to say, $E=mc^2$ where E is energy and *m* is mass: and he remarked that if this were so, then a Hertz oscillator, which sends out electromagnetic energy preponderantly in one direction, should recoil as a gun does when it is fired. In 1904 F. Hasenöhrl[4] (1874-1915) considered a hollow box with perfectly reflecting walls filled with radiation, and found that when it is in motion there is an (*continued to next page*) apparent addition to its mass, of amount $(8/3c^2)$ times the energy possessed by the radiation when the box is at rest: in the following year[1] he corrected this to $(4/3c^2)$ times the energy possessed by the radiation when the box is at rest[2]; that is, he agreed with J. J. Thomson's $E=(3/4)mc^2$ rather than with Poincaré's $E=mc^2$. In 1905 A. Einstein[3] asserted that when a body is losing energy in the form of radiation its mass is diminished approximately (i.e. neglecting quantities of the fourth order) by ($1/c^2$) times the energy lost. He remarked that it is not essential that the energy loss by the body should consist of radiation, and suggested the general conclusion, in agreement with Poincaré, that the mass of a body is a measure of its energy content: if the energy changes by E ergs, the mass changes in the same sense by E/$c^2$ grams. In the following year he claimed[4] that this law is the necessary and sufficient condition that the law of conservation of motion of the centre of gravity should be valid for systems in which electromagnetic as well as mechanical processes are taking place." (We refer the readers to [2] for footnotes and references in the quotation except noting that the Einstein's two references are [29, 30] and that further studies of Fermi (1922), Wilson (1936), von Mosengeil (1907) and Planck (1907) corrected both cases with E=(3/4)$mc^2$ to agree with E=$mc^2$.)

*Further developments in special relativity.* The development of special relativity continued after 1905. Planck in 1906 [31] obtained the relativistic formulas of kinetic energy and momentum of a material particle. Minkowski in 1907 derived the 4-dimensional covariant formulation of the Maxwell's equations together with the 4-dimensional stress-energy tensor of electromagnetic field [32]. We will address more of these developments relevant to the genesis of general relativity.

*Differential geometry and tensor calculus.* In 1854, Riemann [33] founded Riemannian geometry. Metric was the fundamental entity in Riemannian geometry. Christoffel's [34] introduced covariant differentiation. In 1872 Erlangen program, Klein [35] first gave a generalized definition of geometry and cleared indicated the essential nature of a vector under the group of rotations of orthogonal axes in 3-dimensional space. Various authors [36-39] drew attentions to symmetric tensors of rank 2, scalars and tensors of rank 2. From 1887 onwards, Ricci-Curbastro generalized the theory to tensor calculus for transformations in curved space of any dimensions. It became widely known when Ricci (Ricci-Curbastro) and Levi-Civita [40] published their memoir describing it in 1900. These developments greatly facilitated the development of general relativity.

Table 2 lists important historical steps toward synthesis of a new theory of gravity (post Newtonian theory) agreeing with experiment/observation before the genesis of special relativity in 1905.

Table 2. Historical steps toward synthesis of a new theory of gravity (post Newtonian theory) before the genesis of special relativity in 1905.

| Year | Reference | Historical Step |
|---|---|---|
| 1859 | Le Verrier [6] | Discovery of Mercury perihelion advance anomaly |
| 1882 | Newcomb [9] | Improved measurement of the Mercury perihelion advance anomaly |
| 1887 | Michelson and Morley [12] | Michelson-Morley experiment |
| 1889 | Eötvös [14] | Eötvös experiment to test WEP to $10^{-8}$ level |
| 1864 on | See, e.g. Roseveare [8] | The proposals of various new gravity theories |
| 1854-1900 | Riemann [33], Klein [35], Ricci and Levi-Civita [40] | The development of differential geometry and tensor analysis |
| 1887-1904 | Lorentz [20], and various authors | Approximate transformation theory of special relativity |
| 1900-1904 | Poincaré [21-23] | Principle of relativity |
| 1905 | Poincaré [27], Einstein [28] | Exact transformation theory of special relativity |
| 1905 | Einstein [29] | $E = mc^2$ in special relativity |

## 2. The Period of Searching for Directions and New Ingredients: 1905-1910

*The genesis of general relativity can be roughly divided into 3 periods: (i) 1905-1910, the period of searching for directions and ingredients; (ii) 1911-1914, the period of various trial theories; (iii) 1915-1916, the synthesis and consolidation.* In the prelude we have seen that Newton's gravitation theory needs to be replaced. In this section we first discuss some ingredients of it followed by searching for directions and new ingredients towards genesis of a new gravitation theory.

*Ingredients of Newton's theory.* Newton's theory of gravity is an inverse law with active gravitational mass proportional to passive gravitational mass and active gravitational mass also proportional to inertial mass. With appropriate choice of units, the gravitational force $F_{1 \to 2}$ acting on body 2 from body 1 can be written in the form

$$F_{1 \to 2} = G_N \, m_{a1} \, m_{p2} \, \mathbf{n}_{1 \leftarrow 2} / r_{12} = m_{iner2} \, \mathbf{a}_2, \tag{1}$$

where $m_{a1}$ is the active gravitational mass of body 1, $m_{p2}$ the passive gravitational mass of body 2, $\mathbf{n}_{1 \leftarrow 2}$ the unit vector from body 2 to body 1, $r_{12}$ the distance between body 1 and body 2, $m_{iner2}$ the inertial mass of body 2, $\mathbf{a}_2$ the acceleration of body 2, and $G_N$ the universal Newton constant. The Galileo weak equivalence principle dictates the equality of passive gravitational mass and the inertial mass, i.e. $m_p = m_{iner} \equiv m$ while Newton's third law of motion dictates the equality of passive gravitational mass and the active gravitational mass, i.e. $m_p = m_a = m$. Hence (1) becomes

$$F_{1 \to 2} = G_N \, m_1 \, m_2 \, \mathbf{n}_{1 \leftarrow 2} / r_{12} = m_2 \, \mathbf{a}_2. \tag{2}$$

The action is instant. In Newton's original form the theory is an action-at-a-distance theory. In potential theory form, the gravitational potential $\Phi(\mathbf{x}, t)$ for a mass distribution $\rho(\mathbf{x}, t)$ satisfies the Poison equation:

$$\nabla^2 \Phi(\mathbf{x}, t) = 4\pi \, G_N \, \rho(\mathbf{x}, t). \tag{3}$$

The left-hand side of (3) depends on the gravitational field while the right-hand side depends on the gravitating source. In the field approach, to reach a new theory of gravity we may need to replace both the left-hand side and right-hand side.

*Finite velocity of propagation.* It is natural for Poincaré who reached the exact transformation theory in agreement with the principle of relativity to also think about how to reconcile gravity. Poincaré [27, 41] pointed out that for principle of relativity to be true, gravity must be propagated with speed of light, and mentioned gravitational-wave propagating with the speed of light based on Lorentz invariance. He attempted to formulate an action-at-a-distance theory of gravity with finite propagation velocity compatible with principle of relativity, but was unsuccessful.

*All energy must gravitate.* As we mentioned in the last section, Planck [31] obtained the relativistic formulas of kinetic energy and momentum of a material particle in 1906. Since energy is equivalent to mass and has inertia, it must gravitate according to the equivalence of the inertia mass and the gravitational mass which was verified to great precision by Eötvös experiment. Hence, Planck [42] postulated that all energy must gravitate in 1907 and made another step toward a new theory of gravity.

*Einstein equivalence principle.* Einstein [43], in the last part (Principle of Relativity and Gravitation) of his Comprehensive 1907 essay on relativity, proposed the complete physical equivalence of a homogeneous gravitational field to a uniformly accelerated reference system: "We consider two systems of motion, $\Sigma_1$ and $\Sigma_2$. Suppose $\Sigma_1$ is accelerated in the direction of its *X*-axis, and *γ* is the magnitude (constant in time) of this acceleration. Suppose $\Sigma_2$ is at rest, but situated in a homogeneous gravitational field, which imparts to all objects an acceleration −*γ* in the direction of the *X*-axis. As far as we know, the physical laws with respect to $\Sigma_1$ do not differ from those with respect to $\Sigma_2$, this derives from the fact that all bodies are accelerated alike in the gravitational field. We have therefore no reason to suppose in the present state of our experience that the systems $\Sigma_1$ and $\Sigma_2$ differ in any way, and will therefore assume in what follows the complete physical equivalence of the gravitational field and the corresponding acceleration of the reference system."

From this equivalence, Einstein derived clock and energy redshifts in a gravitational field. The reasoning is clear and simple: two observers at different location of the uniform gravitational field can be equivalently considered in an accelerated frame. In the equivalent accelerated frame there are Doppler shift. This gives redshift/blueshift in the gravitational field. When applied to a spacetime region where inhomogeneities of the gravitational field can be neglected, this equivalence dictates the behavior of matter in gravitational field. The postulate of this equivalence is called the Einstein Equivalence Principle (EEP). EEP is the cornerstone of the gravitational coupling of matter and non-gravitational fields in general relativity and in metric theories of gravity. EEP fixes local physics to be special relativistic.

Local physics in Newtonian gravity also observed this equivalence principle formally except here the local physics is Newtonian mechanics, not special relativity (Here the transformation to the accelerated frame is through a non-Galilean transformation. See, e.g. [19] and references therein for details.).

*Four dimensional spacetime formulation and the Minkowski metric*. On 21 December 1907, Minkowski read before the Academy "Die Grundgleichungen für die elektromagnetischen Vorgänge in bewegten Körpern" (The fundamental equations for electromagnetic processes in Moving bodies) [32] (See also [44]). In this paper, Minkowski put Maxwell equations into geometric form in four-dimensional spacetime with Lorentz covariance using Cartesian coordinates *x*, *y*, *z* and imaginary time *it* and numbering them as $x_1 \equiv x$, $x_2 \equiv y$, $x_3 \equiv z$ and $x_4 \equiv it$. Minkowski defined the 4-dim excitation in terms of ***D*** and ***H***, and the 4-dim field strength in terms of ***E*** and ***B***.

Maxwell equations in Minkowski form was soon written in integral form by Hargreaves [45] and devoted a detailed investigation by Bateman [46] and Kottler [47].

In 1909, Bateman [46] worked on the electrodynamic equations. He used time coordinate *t* instead of $x_4$, and studied integral equations and the invariant transformation groups. He considered specifically transformations that leave the invariance of the differential (form) equation:

$$(dx)^2 + (dy)^2 + (dz)^2 - (dt)^2 = 0, \qquad (4)$$

and included conformal transformations in addition to Lorentz transformations, therefore he went one step forward toward general coordinate invariance. He did use more general (indefinite) metric from coordinate transformations in his study of electromagnetic equation.

With the definition $x^1 \equiv x$, $x^2 \equiv y$, $x^3 \equiv z$ and $x^0 \equiv t$, Eq. (4) can be written as

$$(dx^1)^2 + (dx^2)^2 + (dx^3)^2 - (dx^0)^2 = - \eta_{ij} dx^i dx^j = 0, \tag{5}$$

where the Minkowski metric $\eta_{ij}$ is defined as

$$\eta_{kl} = \begin{pmatrix} 1 & 0 & 0 & 0 \\ 0 & -1 & 0 & 0 \\ 0 & 0 & -1 & 0 \\ 0 & 0 & 0 & -1 \end{pmatrix}, \tag{6a}$$

with its inverse $\eta^{kl}$

$$\eta^{kl} = \begin{pmatrix} 1 & 0 & 0 & 0 \\ 0 & -1 & 0 & 0 \\ 0 & 0 & -1 & 0 \\ 0 & 0 & 0 & -1 \end{pmatrix}. \tag{6b}$$

In (5) and this article, we use Einstein convention of summing over repeated indices. Minkowski metric is used in raising and lowering covariant and contravariant indices in special relativity.

With indefinite metric, one has to distinguish covariant and contravariant tensors and indices. Aware of this, one can readily put Maxwell equations into covariant form without using imaginary time. Following Minkowski [32] but using real time coordinate, in terms of Minkowski 4-dim field strength $F_{kl}$ (**E**, **B**) and 4-dim excitation (density) $H^{ij}$ (**D**, **H**)

$$F_{kl} = \begin{pmatrix} 0 & E_1 & E_2 & E_3 \\ -E_1 & 0 & -B_3 & B_2 \\ -E_2 & B_3 & 0 & -B_1 \\ -E_3 & -B_2 & B_1 & 0 \end{pmatrix}, \tag{7a}$$

$$H^{ij} = \begin{pmatrix} 0 & -D_1 & -D_2 & -D_3 \\ D_1 & 0 & -H_3 & H_2 \\ D_2 & H_3 & 0 & -H_1 \\ D_3 & -H_2 & H_1 & 0 \end{pmatrix}. \tag{7b}$$

Maxwell equations can be expressed in Minkowski form as

$$H^{ij}{}_{,j} = -4\pi J^i, \tag{8a}$$
$$e^{ijkl} F_{jk,l} = 0, \tag{8b}$$

where $J^k$ is the charge 4-current density ($\rho$, $\boldsymbol{J}$) and $e^{ijkl}$ the completely anti-symmetric tensor density (Levi-Civita symbol) with $e^{0123} = 1$. "," means partial derivation. In vacuum, the relation of Minkowski 4-dim field strength $F_{kl}$ ($\boldsymbol{E}$, $\boldsymbol{B}$) and 4-dim excitation (density) $H^{ij}$ ($\boldsymbol{D}$, $\boldsymbol{H}$) is

$$H^{ij} = \eta^{ik} \eta^{jl} F_{kl} = \tfrac{1}{2} (\eta^{ik} \eta^{jl} - \eta^{il} \eta^{jk}) F_{kl}, \text{ i.e., } H^{ij} = F^{ij}. \tag{9}$$

*Four dimensional electromagnetic stress-momentum-energy tensor.* In the sme paper, Minkowski [32] derived the 4-dim electromagnetic stress-momentum-energy (or stress-energy or energy-momentum) tensor $T^{(EM)}{}_i{}^j$ of rank 2:

$$T^{(EM)}{}_i{}^j = (1/16\pi)\, \delta_i^j F_{kl} F^{kl} - (1/4\pi) F_{il} F^{jl}, \tag{10}$$

with

$$T^{(EM)}{}_0{}^0 = (1/8\pi)\, [(\boldsymbol{E})^2 + (\boldsymbol{B})^2], \tag{11}$$

the electromagnetic energy density discovered by W. Thomson (Kelvin) in 1983;

$$T^{(EM)}{}_0{}^\mu = -(1/4\pi) F_{0\nu} F^{\mu\nu} = (1/4\pi) (\boldsymbol{E} \times \boldsymbol{B})^\mu, \tag{12}$$

($1/c$) times the electromagnetic energy flux discovered by Pointing and Heaviside in 1884;

$$T^{(EM)}{}_\mu{}^0 = -(1/4\pi) F_{\mu\nu} F^{0\nu} = (1/4\pi) (\boldsymbol{E} \times \boldsymbol{B})_\mu (= -(1/4\pi) (\boldsymbol{E} \times \boldsymbol{B})^\mu), \tag{13}$$

($-c$) times the electromagnetic momentum density discovered by J. J. Thomson in 1893;

$$T^{(EM)}{}_\mu{}^\nu = (1/16\pi)\, \delta_\mu^\nu F_{kl} F^{kl} - (1/4\pi) F_{\mu\alpha} F^{\nu\alpha}$$
$$= (1/8\pi)\, \{\delta_\mu^\nu [(\boldsymbol{E})^2 + (\boldsymbol{B})^2] - \eta_{\mu\alpha} [(\boldsymbol{E})^\alpha (\boldsymbol{B})^\nu + (\boldsymbol{B})^\alpha (\boldsymbol{E})^\nu]\}, \tag{14}$$

the electromagnetic stress discovered by Maxwell in 1973. Here we use Greek indices to run from 1 to 3.

The importance of constructing the 4-dim electromagnetic stress-momentum-energy tensor is that it was the first 4-dim stress-momentum-energy tensor ever constructed. For electromagnetic energy to gravitate, it should enter the right-hand side of the new covariant (3). However, electromagnetic energy is only the (0, 0) component of 4-dim stress-momentum-energy tensor, other components should enter the right-hand also to make it covariant.

*Directions and new ingredients.* During 1905-1910, directions and new ingredients were formed for a new theory of gravity. We had finite propagation velocity, all energy gravitating, EEP, spacetime formulation of special relativity, indefinite metric, and 4-dim covariant electromagnetic stress-momentum-energy (stress-energy) tensor. Two crucial steps are (i) the generalization of the principle of relativity to include situation in gravity, i.e. the Einstein Equivalence Principle; and (ii) the spacetime formulation of the (special) relativity theory using Minkowski metric and its generalization to the general concept of indefinite spacetime metric. EEP means the local physics is special relativistic. Then one can ask what gravity is. It must be how various local physics are connected. We have special relativity from locality to locality and gravity describes how they are connected. (A mathematical natural description is a four-dimensional base manifold with special relativity as fibre attached to each (world) point in the base manifold, and gravity is the connection bundle or the metric which induces the connection bundle.) Although this logic seems compelling, the full metric as dynamical gravitational entity was not used until 1913. A test of EEP was derived by Einstein: the gravitational redshift. It has been an important test of relativistic gravity which people try to improve the accuracy constantly.

### 3. The Period of Various Trial Theories: 1911-1914

*Basic formulas of (pseudo-)Riemannian geometry.* Here we summarize some basic formulas used in developing a new theory of gravity for straightening out the convention and notation. First, a (pseudo)-Riemannian manifold is endowed with a metric $g_{ij}$. The metric $g_{ij}$ is related to the line element $ds$ as:

$$ds^2 = g_{ij}\, dx^i\, dx^j. \tag{15}$$

If the metric $g_{ij}$ is positive definite, the geometry is Riemannian. If the metric $g_{ij}$ is indefinite, the geometry is pseudo-Riemannian. $g^{ij}$ is the matrix inverse of $g_{ij}$ and they are used to raise and lower covariant and contravariant indices. For our case, the geometry is pseudo-Riemannian. We use the MTW [48] conventions with signature −2; this is also the convention used in [49]. Latin indices run from 0 to 3; Greek indices run from 1 to 3. The Christoffel connection $\Gamma^i{}_{jk}$ of the metric is given by

$$\Gamma^i{}_{jk} = \tfrac{1}{2}\, g^{il}\, (g_{lj,k} + g_{lk,j} - g_{jk,l}). \tag{16}$$

With Christoffel connection, one can define covariant derivative. The Riemannian curvature tensor $R^i{}_{jkl}$, the Ricci curvature tensor $R_{jl}$, the scalar curvature $R$ and the Einstein tensor $G_{jl}$ are defined as:

$$R^i{}_{jkl} = \Gamma^i{}_{jl,k} - \Gamma^i{}_{jk,l} + \Gamma^i{}_{km}\Gamma^m{}_{lj} - \Gamma^i{}_{lm}\Gamma^m{}_{jk};\ R_{kj} = R^i{}_{jil};\ R = g^{jl}R_{jl};\ G_{jl} = R_{jl} - (1/2)g_{jl}R. \qquad (17)$$

*Gravitational deflection of light and EEP.* Extending his work on gravitational redshift, Einstein [50] derived light deflection in gravitational field using EEP in 1911. He argued that since light is a form of energy, light must gravitate and the velocity of light must depend on the gravitational potential. He obtained that light passing through the limb of the Sun would be gravitationally deflected by 0.83 arc sec. This is very close to the value 0.84 arc sec derived by Soldner [51] in 1801 assuming that light is corpuscular in Newtonian theory of gravitation. This prediction was half the value of general relativity. Before 1919, there were 4 expeditions intent to measure the gravitational deflection of starlight (in 1912, 1914, 1916 and 1918); because of bad weather or war, the first 3 expeditions failed to obtain any results, the results of 1918 expedition was never published [52]. In 1919, the observation of gravitational deflection of light passing near the Sun during a solar eclipse [53] confirmed the relativistic deflection of light and made general relativity famous and popular.

*Stress-energy tensor.* In 1911, Laue [54] extended Minkowski's method of constructing electromagnetic stress-energy tensor to stressed bodies, dust and relativistic fluids.

*Gravity theories with 'variable velocity of light' and scalar theories of gravity.* Accepting that the velocity of light depends on gravitational potential, Abraham [55] postulated that the negative gradient indicates the direction of gravitational force and worked out a theory of gravity. Einstein [56] worked out a somewhat different theory. These gravity theories with 'variable velocity of light' led to the proposals of conformally flat scalar theories of Nordström [57-59].

The equation corresponding to Eq. (3) in Newtonian theory for electromagnetism is

$$(1/c^2)(\partial^2 A_i/\partial t^2) - \nabla^2 A_i = A_{i,}{}^j = 4\pi J_i, \qquad (18)$$

with gauge condition

$$A_{i,}{}^i = 0. \qquad (19)$$

Here $A_i$ is the electromagnetic 4-potential guaranteed locally by (8b) such that $F_{ij} = A_{j,i} - A_{i,j}$. To incorporate the finite propagation speed with light velocity into the gravitation field equation, one could just replace (3) with

$$(1/c^2)(\partial^2 \Phi^*/\partial t^2) - \nabla^2 \Phi^* = \eta^{ij}\Phi^*_{,ij} = -4\pi G_N \rho^*(\boldsymbol{x}, t), \qquad (20)$$

where $\Phi^*(\boldsymbol{x}, t)$ is a new gravitational field entity. $\Phi^*$ could be a scalar field, a vector field or a tensor field or some combination of them. If $\Phi^*$ is a scalar field, $\rho^*$ must be a scalar; in the weak field and slow motion limit, one must be able to approximate $\Phi^*$ and $\rho^*$ by $\Phi$ and $\rho$. Let us illustrate with Einstein's theory with 'variable velocity of light'.

In the original formulation of Einstein [56], the equation of motion for particles was derived from the variational principle

$$\delta \int ds = 0, \tag{21}$$

where

$$ds^2 = c^2 dt^2 - dx^2 - dy^2 - dz^2, \tag{22}$$

and where $c$ is a scalar function which Einstein regarded as the velocity of light in the metric (22). Einstein postulated that $c$ depends on the scalar field $\varphi$ in the following way:

$$c^2 = c_0^2 - 2\varphi, \tag{23}$$

and that $\varphi$ is generated by $\rho^*$ through the wave equation

$$(1/c^2)(\partial^2\varphi/\partial t^2) - \nabla^2\varphi(\boldsymbol{x}, t) = -4\pi\, G_N\, \rho^*(\boldsymbol{x}, t). \tag{24}$$

By choosing suitable units, we can set $c_0 = 1$; and by postulating that Einstein's $ds^2$ is the "physical metric," we can bring the theory into the form:

$$ds^2 = (c_0^2 - 2\varphi)\, dt^2 - dx^2 - dy^2 - dz^2, \tag{25a}$$
$$\eta^{ij}\varphi_{,ij} = 4\pi\, G_N\, \rho^*. \tag{25b}$$

Note that in (25b) as well in (20), we have the *a priori* (non-dynamical) geometric element $\eta^{ij}$ to make the equation fully coordinate covariant. More precisely, equation (25a) also contains *a priori* geometric elements – a flat-space metric and a time direction. This makes the theory a stratified theory with conformally flat space slices. For more detailed discussions, see [60, 61].

The physical metric can always be transformed locally into the Lorentz form

$$ds^2 = c_0^2 d\underline{t}^2 - d\underline{x}^2 - d\underline{y}^2 - d\underline{z}^2, \tag{26}$$

where $d\underline{t}$ is the proper time interval and $dl = (d\underline{x}^2 + d\underline{y}^2 + d\underline{z}^2)^{1/2}$, the proper-length element. Since light trajectories all lie on null cones of this metric, the velocity of light as measured using the physical metric is always $c_0$ -- as it must be for any theory that satisfy the Einstein equivalence principle.

This theory did not agree with the Mercury perihelion advance observation. However, it led to the conformally flat theories of Nordström [57-59].

The field equations of Nordström's second theory [58, 59] can be written as

$$C_{ijkl} = 0, \tag{27}$$
$$R = 24\pi(G_N/c^4)T, \tag{28}$$

where $C_{ijkl}$ is the Weyl conformal tensor and $R$ is the curvature scalar both constructed from the metric $g_{ij}$. $T$ is the trace contraction of stress-energy tensor. The field equations (27) and (28) are geometric and make no reference to any gravitational fields except the physical metric $g_{ij}$. However, they guarantee the existence of a flat spacetime metric $\eta_{ij}$ (prior geometry in the language of [48]) and a scalar field related to $g_{ij}$ by

$$g_{ij} = \varphi^2 \eta^{ij}; \tag{29}$$

and they allow $\varphi$ to be calculated from the variational principle

$$\delta \int [L_\text{I} - (1/3)\, R\, (-g)^{1/2}] d^4x = 0, \tag{30}$$

where $g = \det(g_{ij})$ and $L_\text{I}$ is the interaction Lagrangian density of matter with gravity (see [60] for more details). Expressed in terms of $\varphi$, the field equation (28) becomes

$$\eta^{ij}\varphi_{,ij} = -(4\pi G_\text{N}/c^4)\, T\, \varphi^3 \tag{31a}$$

or

$$\eta^{ij}\varphi_{,ij}\varphi^{-1} = -(4\pi G_\text{N}/c^4)\, T_\text{flat}. \tag{31b}$$

Equation (31) is Nordström's original field equation [58, 59], while equation (28) is the Einstein-Fokker version [62]. This second Nordström theory did not agree with the Mercury perihelion advance observation either. From (14), we notice that the trace contraction of electromagnetic tensor vanishes. Therefore, in this theory the electromagnetic energy does not contribute to the generation of gravitational field. Neither the gravitational field gives light deflection. Somehow Nature did not choose this way. Nature chose to make the whole metric dynamic.

*Tensor theory of gravity.* In 1913, Einstein and Grossmann turned into tensor theory of gravity making full use of the metric. They tried to incorporate all the ingredients discussed in the last section into their "Entwurf (outline)" theory [63] and proposed the following equation using the metric $g_{ij}$ as dynamical entity for the gravitational field:

*Part of Ricci tensor $R_{ij} \propto T_{ij}$.* (32)

Since the left-hand side did not contain all the terms of the Ricci tensor, it is not covariant. In 1913 Besso and Einstein [64] worked out a Mercury perihelion advance formula in the "Einstein-Grossmann Entwurf" theory [63], but the calculation contained an error and the result did not agree with the Mercury perihelion advance observation. Nevertheless, the "Entwurf" theory is an important landmark in the genesis of general relativity.

Einstein became versed at differential geometry and tensor analysis in 1914. A quote of Einstein's October 1914 writing on "The formal foundation of the general theory of relativity" [65] showed the situation: in the abstract "In recent years I have worked, in part with my friend Grossmann, on a generalization of the theory of relativity. During these investigations, a kaleidoscopic mixture of postulates from physics and mathematics has been introduced and used as heuristical tools; as a consequence it is not easy to see through and characterize the theory from a formal point of view, that is, only based upon these papers. The primary objective of the present paper is to close this gap. In particular, it has been possible to obtain the equations of the gravitational field in a purely covariance-theoretical manner (section D). I also tried to give simple derivations of the basic laws of absolute differential calculus – in part, they are probably new ones (Section B) – in order to allow the reader to get a complete grasp of the theory without having to read other, purely mathematical tracts. As an illustration of the mathematical methods, I derived the Eulerian equations of hydrodynamics and the field equations of the electrodynamics of moving bodies (section C). Section E shows that Newton's theory of gravitation follows from the general theory as an approximation. The mist elementary features of the present theory are also derived inasfar as they are characteristic of a Newtonian (static) gravitational field (curvature of light rays, shift of spectral lines)."

In the 1911-1914 period, various trial theories, based largely on the ingredients and directions set in the previous period 1905-1910, emerged and led step by step towards the synthesis of general relativity

## 4. The Synthesis and Consolidation: 1915-1916

*Einstein's big step*. Continued along the direction set in the "Entwurf" theory, Einstein [66, 67] reached the following equation for general relativity in 1915:

$$R_{ij} \propto T_{ij}. \tag{33}$$

Subsequently, Einstein [68] corrected an error made in his collaboration with Besso of 1913 [64] and obtained a value of Mercury perihelion advance from his new equation (33) in agreement with the observation [9]. Apparently, this correct calculation played a significant role in the final genesis of general relativity. The divergence of $T_{ij}$ vanishes. However, the divergence of $R_{ij}$ does not vanish unless $T$ vanishes or is constant. Since the trace $T^{(EM)}$ of electromagnetic stress-energy tensor does vanish, Einstein [67] argued that:

"One now has to remember that by our knowledge "matter" is not to be perceived as something primitively given or physically plain. There even are those, and not just a few, who hope to reduce matter to purely electrodynamic processes, which of course would have to be done in a theory more completed than Maxwell's electrodynamics. Now let us just assume that in such completed electrodynamics scalar of the energy tensor also would vanish! Would the result, shown above, prove that matter cannot be constructed in this theory? I think I can answer this question in the negative, because it might very well be that in "matter," to which the previous expression relates, gravitational fields do form an important constituent. In that case, $\Sigma\ T_\mu{}^\mu$ can appear positive for the entire structure while in reality only $\Sigma\ (T_\mu{}^\mu + t_\mu{}^\mu)$ is positive and $\Sigma\ T_\mu{}^\mu$ vanishes every-where. *In the following we assume the conditions $\Sigma\ T_\mu{}^\mu = 0$ really to be generally true*."

*Hilbert variational principle*. Shortly after Einstein obtained (33), Hilbert [69] proposed the variational principle for gravitational field:

$$\delta \int [L_I^{(EM)} + L^{(GRAV)}] \, d^4x = 0, \qquad (34)$$

with the Lagrangian densities $L_I^{(EM)}$ and $L^{(GRAV)}$ given by

$$L_I^{(EM)} \propto F_{kl} F^{kl} (-g)^{1/2}, \qquad (35a)$$

$$L^{(GRAV)} \propto R (-g)^{1/2}. \qquad (35b)$$

The variation of integral of (35a) plus the Lagrangian density term of electromagnetic 4-current interaction with electromagnetic 4-potential gives the Maxwell equations. The variation of the integral of (35a) with respect to the metric gives the electromagnetic stress-energy tensor. The variation of (35b) together with (35a) with respect to the metric would give:

$$R_{ij} - \tfrac{1}{2} g_{ij} R \propto T^{(EM)}_{ij}. \qquad (36a)$$

With more general Lagrangian $L_I$ such as Mie's Lagrangian instead of $L_I^{(EM)}$, the variation of its integral and Hilbert's gravitational integral with respect to the metric would give

$$R_{ij} - \tfrac{1}{2} g_{ij} R \propto T_{ij}. \qquad (36b)$$

Here $T_{ij}$ is given using Hilbert's variation-with-respect-to-the-metric definition.

*Einstein equation.* After Hilbert's work [69], Einstein [70] soon corrected his field equation (33) on November 25, 1915 to

$$R_{ij} \propto T^{(EM)}_{ij} - \tfrac{1}{2} g_{ij} T. \qquad (37)$$

Equation (36b), or Equation (36a) with $T^{(EM)}_{ij}$ replaced by $T_{ij}$, i.e.

$$R_{ij} - \tfrac{1}{2} g_{ij} R \propto T_{ij}, \qquad (38)$$

and equation (37) are equivalent to each other: by taking a trace contraction of either equation, one has

$$R \propto -T, \qquad (39)$$

and from (39) the equivalence becomes clear. Since variation principle became common, the Einstein equation is normally written in the form (38) nowadays. With the proportional constant inserted, the Einstein equation is

$$G_{ij} = R_{ij} - \tfrac{1}{2}\, g_{ij}\, R = 8\pi (G_N/c^4)\, T_{ij}. \tag{40}$$

In 1916, Einstein [71] wrote a foundational paper on general relativity. In the same year, Einstein performed a linear approximation in the weak field and obtained the quadrupole radiation formula [72]; major errors were corrected in his 1918 paper [73] while a factor of 2 was corrected by Eddington [74]. (For later controversial issues on gravitational wave and the quadrupole formula, see e.g. Ref.'s [75, 76].) Einstein thought that quantum effects must modify general relativity in his first paper on linear approximation and gravitational waves [72] although he switched to a different point of view working on the unification of electromagnetism and gravitation in the 1930s. The merging of general relativity and quantum theory is an important issue. For a brief history of ideas and prospects, see e.g. Ref. [77].

In 1916, Schwarzschild discovered an exact spherical solution (Schwarzschild solution) of Einstein equation [78, 79].

In 1917, Einstein [80] postulated the cosmological principle, applied general relativity to cosmology and proposed the cosmological constant; de Sitter [81-84] followed in the same year with an inflationary solution to cosmology. Ever since the genesis of general relativity, it went hand-in-hand with the development of cosmology. For a brief history of this connection and mutual development, see e.g. Ref. [85]; for recent reviews on various topics in cosmology, see e.g. Ref. [86-91].

In 1918, Lense and Thirring [92] discovered the frame-dragging effect in general relativity.

After one hundred years of developments, general relativity becomes indispensable in precision measurement, astrophysics, cosmology and theoretical physics. The first direct detections of gravitational waves [93, 94] in the centennial of the genesis of general relativity truly celebrate this occasion.

The route to general relativity is indeed guided by covariance. However, when general relativity is reached and covariance is fully understood, the principle of covariance could accommodate various things, scalars, vectors, a priori objects etc. and various theories of gravity. It is probably a *minimax principle* that worked in nature: *When an entity is needed, it should saturate its maximal capacity*.

In Table 3, we list historical steps in the genesis of general relativity discussed in the last section and this section.

## 5. Epilogue

The study of histories gives inspiration. The study of the genesis of general relativity is clearly so. It is fortunate that most of the records are intact and the step-to-step development are transparent. We hope that this short exposition presents the flavor and some insights of the development. The genesis of general relativity was a community effort with Einstein clearly dominated the scene. Knowledge accrues gradually most of the time. Cognition sometimes comes in somewhat bigger steps. The cognition that energy must gravitates and that the Einstein equivalence principle must be valid are such examples. Initial cognition needs consolidation and development. Einstein equivalence principle indicates that local physics must be special relativistic. Minkowski's spacetime formulation indicates that local physics must have an indefinite metric. To take metric as basic entity for gravitation took a few years through studying gravitational redshift, gravitational light deflection, theories of gravity with "variable velocity of light" and more scalar theories of gravity. Eventually, metric as a full dynamic entity for gravitation emerged in 1913. It took a couple of years to master this approach for leading to the genesis of general relativity. During different phases of genesis, the first discovered general relativistic effect – the Mercury perihelion advance anomaly played a key role.

Table 3. Historical steps in the genesis of general relativity since the genesis of special relativity.

| Year | Reference | Historical Step |
| --- | --- | --- |
| 1905-06 | Poincaré [27, 41] | Attempt to formulate an action at a distance theory of gravity with finite propagation velocity compatible with principle of relativity |
| 1907 | Planck [42] | All energy must gravitate. |
| 1907-08 | Einstein [43] | Generalized principle of relativity (Einstein Equivalence Principle [EEP]) and the prediction of gravitational redshift |
| 1907-08 | Minkowski [32, 44] | Covariant spacetime formulation of electromagnetism and the derivation of 4-dim electromagnetic stress-energy tensor |
| 1909-10 | Bateman [46] | Introducing indefinite spacetime metric |
| 1911 | Einstein [50] | Using EEP to derive deflection of light in gravitational field |
| 1911 | Laue [54] | Stress-energy tensor of matter |
| 1911-12 | Abraham [55], Einstein [56] | Theories with 'variable velocity of light' |
| 1912 | Nordström [57] | Norström's first theory |
| 1913 | Nordstrom [58, 59] | Norström's second theory |
| 1913 | Einstein and Grossman [63] | "Entwurf (Outline)" theory |
| 1914 | Einstein and Fokker [62] | Covariant formulation of Norström's second theory |
| 1914 | Einstein [65] | Einstein versed at covariant formulation |
| 1915 | Einstein [66, 67] | Source restricted Einstein equation |
| 1915 | Hilbert [69] | Hilbert variational principle |
| 1915 | Einstein [70] | Einstein equation |
| 1916 | Einstein [71] | Einstein's foundation paper of general relativity |
| 1916 | Einstein [72] | Approximate solution and gravitational waves |
| 1916 | Schwarzschild [77] | Exact spherical solutions of Einstein equation |
| 1917 | Einstein [80] | Cosmological principle, cosmology and cosmological constant |
| 1917 | de Sitter [80-84] | de Sitter inflationary solution (cosmology) |
| 1918 | Einstein [73, 74] | Quadrapole radiation formula |
| 1918 | Lense-Thirring [92] | Lense-Thirring gravitomagnetic effect |

**Acknowledgements**

I would like to thank Science and Technology Commission of Shanghai Municipality (STCSM-14140502500) and Ministry of Science and Technology of China (MOST-2013YQ150829, MOST-2016YFF0101900) for supporting this work in part.